\documentclass[%
 reprint,
 amsmath,amssymb,
 aps,
]{revtex4-1}
\usepackage{graphicx}
\usepackage{dcolumn}
\usepackage{bm}
\usepackage[normalem]{ulem}	

\begin{document}

\preprint{APS/123-QED}

\title{Energy Density, Temperature and Entropy Dynamics in Perturbative Reheating}
\author{Alessandro Di Marco}
\email{alessandro.di.marco@roma2.infn.it}
\author{Giancarlo De Gasperis}
\author{Gianfranco Pradisi}
\author{Paolo Cabella}

\affiliation{%
University of Rome Tor Vergata, Via della Ricerca Scientifica 1, 00133 Roma, Italy\\
INFN Sezione di Roma “Tor Vergata”, Via della Ricerca Scientifica 1, 00133 Roma, Italy}%


\begin{abstract}

We discuss the perturbative decay of the energy density of a non standard inflaton field $\rho_{\phi}$
and the corresponding creation of the energy density of the relativistic fields $\rho_r$ at the end of inflation, in the perfect fluid description,
refining some concepts and providing some new computations.
In particular, the process is characterized by two fundamental time scales.
The first one, $t_\text{max}$, occurs when the energy density $\rho_r$  reaches its largest value, slightly after the beginning of the reheating phase. 
The second one, $t_\text{reh}$, is the time in which the reheating is completely realized and the thermalization is attained.
By assuming a non-instantaneous reheating phase, we are able to derive the energy densities and the temperatures of the produced relativistic bath at $t_\text{max}$ and $t_\text{reh}$, as well as the value of the corresponding horizon entropy $S_\text{hor}$, for an Equation-of-State (EoS) parameter $w\ne 0$.

\end{abstract}

\pacs{Valid PACS appear here}
\maketitle


\section{Introduction}

The inflationary mechanism \cite{1,2,3,4,5,6} offers an elegant way to solve the old puzzles of the Big Bang cosmology, i.e. 
the flatness problem, the horizon problem and the monopoles problem, providing at the same time an elegant explanation of the origin of primordial metric fluctuations, that lead to the matter inhomogeneities responsible for the large scale structures 
and to the cosmic microwave background (CMB) anisotropies.
Moreover, inflation is also important because it provides a valuable creation process for matter and radiation currently observed in the Universe. 
This process is often known as the reheating scenario.
In the simplest case, reheating is realized through the introduction of a simple slow-rolling scalar field \cite{6},
the inflaton, that after exploring an almost flat region of the corresponding effective potential driving inflation, falls towards the minimum, oscillates around it and decays giving rise at least to the Standard Model degrees of freedom.
The matter creation process, that depends on the inflationary model, must be driven both by  
perturbative \cite{7,8,9,10} and non-perturbative \cite{11,12,13,14,15,16,17,18} mechanisms and can be modeled 
using many different approaches \cite{19,20,21,22,23,24}.
The whole creation process can be very complicated, 
especially at the first preheating stage, where an exponentially increasing
occupation number is needed, requiring non-perturbative out-of-equilibrium effects. However, at least where thermalization is already important, one can always simplify the picture by using a perfect fluid effective description to model 
the perturbative decreasing of the energy density $\rho_{\phi}$ stored in the inflaton field, and the corresponding increasing 
of the energy density $\rho_r$ of the produced fields leading to the radiation-dominated Universe.  
In this paper, we use this very standard setting to derive the dependence of the energy densities from the physical parameters
at the peak and at the end of the reheating phase.  
The paper is organised as follows. In Section II, we review the general properties of the inflationary dynamics in the perfect fluid approximation, introducing the energy densities and addressing the Cauchy problem related to the relativistic matter production during the reheating phase.  
In Section III, we get the standard solutions of the problem with $w=0$ and outline the difference between the top energy scale reached by the relativistic Standard Model particles 
and the energy scale at the end of the reheating process, when the inflaton disappears from the cosmological particle spectrum.
In Section IV, we introduce a clear definition of the involved entropy and 
in Section V, finally, we generalize the process to the case of a generic fluid with an Equation-of-State (EOS) parameter varying in the range $-1/3<w<1/3$. 
The scales $T_{max}$ and $T_{reh}$ increase differently with $w$, while
the entropy increases in time almost linearly with $w$ for $w > 0$ and more
than quadratically for $w < 0$.  
In this manuscript, we use the particle natural units $\hbar=c=1$ and $M_{p}^2=1/8 \pi G$ will be the (squared of) the reduced Planck mass.

\section{The Cauchy problem for inflaton and radiation energy densities}

In the simplest version of the inflationary scenario, the early Universe should have undergone a phase
driven by a neutral scalar field minimally coupled to gravity. The related action can be written in the form 
\begin{eqnarray}
S\sim \int d^4 x\sqrt{-g} \left(-\frac{1}{2}\partial_{\mu}\phi\partial^{\mu}\phi - V(\phi) \right) \ ,
\end{eqnarray}
using for the space-time a Friedmann-Lemaitre-Robertson-Walker (FLRW) background metric
\begin{eqnarray}
ds^2=-dt^2 + a^2(t)\left[dr^2 + r^2\left(d\theta^2+\sin^2\theta d\varphi^2\right)\right],
\end{eqnarray}
where $t$ is the cosmic time, $(r,\theta,\varphi)$ are the comoving coordinates and $a(t)$ is the dimensionless cosmic scale factor.
The scalar field, the inflaton, is expected to be a weakly self-coupled field equipped with an effective (inflationary) potential $V(\phi)$, characterized by the almost flat slow-roll region that plays the role of a false vacuum, and by a fundamental (true) vacuum state.
The Einstein equations related to the scalar matter result in the Friedmann equation
\begin{eqnarray}\label{eqn: friedmann eq}
H^2&=&\frac{1}{3M_p^2}\rho_{\phi}
\end{eqnarray}
where $H=\dot{a}/a$ is the Hubble rate. 
From the energy-momentum tensor of the scalar field, in analogy with a perfect fluid, one may introduce the energy density $\rho_{\phi}$ and the pressure $p_{\phi}$ 
of the inflaton field as
\begin{eqnarray}\label{eqn: energy density and pressure}
\rho_{\phi}(t)=\frac{1}{2}\dot{\phi}^2 + V(\phi),\quad p_{\phi}(t)=\frac{1}{2}\dot{\phi}^2 - V(\phi) \ ,
\end{eqnarray}
while the equation of motion is given by
\begin{eqnarray}\label{eqn: inflaton eq}
\ddot{\phi}+3H\dot\phi+V'(\phi)=0 \ .
\end{eqnarray}
The inflationary mechanism begins as the inflaton field explores the almost flat region of the scalar potential $V(\phi)\sim M^4_\text{inf}$.
In this phase, the potential term is dominant over the kinetic energy. 
The inflationary phase ends once the inflaton field 
``travels a distance" $\Delta\phi$ \cite{25} to reach the edge of the almost flat region, at some $\phi=\phi_\text{end}$, 
where the slow roll conditions start to be violated.  
The emerging Universe is cold, with energy densities of the (hypothetical) pre-existing fields as well as the pre-inflationary entropy density
suppressed by the almost de Sitter-like expansion.
The inflaton rolls down a potential becoming steeper, so that the kinetic energy density, no-longer negligible, gradually  becomes an important contribution to the field energy density $\rho_{\phi}$.  
As a consequence, the inflaton falls around the minimum (the vacuum state), acquires a mass $m^2=V''(\phi_0)$ and starts to oscillate with a frequency $\omega=m$, naturally larger than the Hubble rate $H$, $m\gg H$.  
Therefore, the period of an oscillation cycle, $t_{\phi}$, is much shorter than an Hubble time scale $t_{\phi}\sim m^{-1}\ll H^{-1}$.  
The system can be interpreted as a condensate of a large number of heavy scalar particles of mass $m$ and zero momentum (in the simplest description with a particle density $n_{\phi}=\rho_{\phi}/m$) that decay, after few oscillations, into relativistic matter.  
Although  the expansion of the Universe induces a damping in the scalar field oscillations,
the coupling of the inflaton to the particles of (an extension of) the Standard Model (SM) can still make the decay very efficient and much larger than opposite effects.  The creation process giving rise to ultra-relativistic particles can be quick or slow and, more importantly, 
it proceeds both via perturbative mechanisms
driven by the decay rate of the interaction \cite{7,8,9,10}, 
and/or via non-perturbative mechanisms, by parametric amplification 
(see \cite{11,12,13,14,15} for some important historical contributions and \cite{16,17,18} for comprehensive reviews).
Thermalization due to scattering among the produced particles leads to the end of reheating and the beginning of the radiation dominance (RD) epoch of Standard Big Bang Cosmology, say at $t_\text{reh}$.  
A simple way to model the quantum particle production in the reheating phase can be introduced
by an additional friction term in the equation of motion of the inflaton field by assuming 
a fast preheating phase where the system is out of equilibrium \cite{7}. We can get an idea about how the total energy density $\rho_{\phi}$ carried by the inflaton field is converted into the energy density $\rho_{r}$ of the radiation products thanks to the perfect fluid formalism.
To this end, we follow a prescription due to Turner in \cite{10}.
In an ``average'' description of the dynamics when thermalization is already important, a proper modified version of the equation of motion of the inflaton field is
\begin{eqnarray}
\ddot{\phi}+3H\dot{\phi}+\Gamma\dot{\phi}+V'(\phi)=0 \ ,
\end{eqnarray}
where the constant $\Gamma$ is mimicking the decay rate of the inflaton
field related to the energy transfer of the scalar boson to the SM particles. By multiplying the equation for $\dot{\phi}$, using  Eqs.($\ref{eqn: energy density and pressure}$) and averaging over an oscillation period one finds
\begin{eqnarray}\label{eqn: form1}
\big \langle\dot{\rho}_{\phi}\big \rangle_{t_{\phi}} = -\big \langle(3H+\Gamma)(\rho_{\phi}+p_{\phi})\big \rangle_{t_{\phi}}.
\end{eqnarray}
The Hubble rate and the decay rate can be taken to be almost constant during the short oscillation cycle $t_{\phi}$ and can be factorized from the average. In addition, since $\rho_{\phi}$ and $p_{\phi}$ are periodic functions, one can define the average of their sum as 
\begin{eqnarray}\label{eqn: sum}
\langle \rho_{\phi}+p_{\phi}\rangle _{t_{\phi}}=\gamma\langle\rho_{\phi}\rangle_{t_{\phi}},
\end{eqnarray} 
where the parameter $\gamma$ is naturally defined as
\begin{eqnarray}\label{eqn: gamma}
\gamma= \langle 1 + p_{\phi}/\rho_{\phi} \rangle_{t_{\phi}}.
\end{eqnarray}
Omitting from now on the brackets, the equation for the evolution of the (average) energy density takes the final form
\begin{eqnarray}\label{eqn: form2}
\dot{\rho}_{\phi}  = -(3H+\Gamma)\gamma \rho_{\phi}.
\end{eqnarray} 
The last step consists of specifying the form of the parameter $\gamma$.
In principle, it is given by
\begin{eqnarray}
\gamma=\frac{1}{t_{\phi}}\int_0^{t_{\phi}} dt  \left( 1+\frac{p_{\phi}}{\rho_{\phi}}\right).
\end{eqnarray}
However, this form of the parameter is quite difficult to handle because it involves an integrand 
that simultaneously depends on time and on the field values.
Anyway, it is possible to show that it can be written as \cite{10}
\begin{eqnarray}
\gamma= 2\frac{I_1(0,\phi_\text{max})}{I_2(0,\phi_\text{max})},
\end{eqnarray}
where
\begin{eqnarray}
I_1(0,\phi_\text{max})=\int_0^{\phi_\text{max}} d\phi\left( 1 - \frac{V(\phi)}{V_\text{max}}\right)^{1/2}
\end{eqnarray}
and
\begin{eqnarray}
I_2(0,\phi_\text{max})=\int_0^{\phi_\text{max}} d\phi\left( 1 - \frac{V(\phi)}{V_\text{max}}\right)^{-1/2} \ ,
\end{eqnarray}
being $V_\text{max}$ the potential energy related to the maximum value of the oscillation of the field $\phi$. 
The parameter $\gamma$ depends on the form of the potential around the minimum.  
For instance, by considering the simple case based on a static inflaton potential of the form $V(\phi)\sim\phi^n$, 
the parameter $\gamma$ can be expressed as the ratio 
\begin{eqnarray}
\gamma= 2 \ \frac{\beta(x,y+1)}{\beta(x,y)}, \quad x=\frac{1}{n}, \quad y=\frac{1}{2},
\end{eqnarray}
where $\beta$ is the Euler beta function.
Consequently, we have 
\begin{eqnarray}
\gamma=\frac{2n}{n+2}
\end{eqnarray}
and the mean value of the equation-of-state (EoS) parameter $w=\gamma-1$ is \cite{26}
\begin{equation}\label{eqn: eos}
w=\frac{n-2}{n+2} \ .
\end{equation}
Moreover, the inflaton potential can be expanded around the minimum as
\begin{eqnarray}
V(\phi)\sim\frac{1}{2}m^2(\phi - \phi_0)^2 + \sum_{n>2} \frac{\lambda^{(n)}}{n!}(\phi-\phi_0)^n
\end{eqnarray}
in the assumption $V(\phi_0)=0$ to avoid a pure cosmological constant contribution.  
Close to the minimum the parabolic term is dominant giving back $\gamma=1$ and consequently $w=0$.
The evolution of the inflaton energy density is so given by
\begin{eqnarray}\label{eqn: inflaton motion}
\dot{\rho}_{\phi} + 3H\rho_{\phi} = - \Gamma\rho_{\phi}
\end{eqnarray} 
namely the familiar equation of a perfect pressure-less fluid (non relativistic matter) with a non zero, negative source term $-\Gamma\rho_{\phi}$ responsible for the decay of the inflaton in other particles.
Eq. \eqref{eqn: inflaton motion} can be rewritten in the form
\begin{eqnarray}
\frac{d (a^3\rho_{\phi})}{dt}=-a^3\Gamma\rho_{\phi}
\end{eqnarray}
that shows how the mean energy density (per comoving volume) $a^3\rho_{\phi}$ is a monotonically decreasing function, being its time variation always negative.  As a consequence, also the number density, proportional to $\rho_{\phi}$, obeys the same equation
\begin{eqnarray}
\frac{d (a^3 n_{\phi})}{dt}=-a^3\Gamma n_{\phi} , 
\end{eqnarray}
and results exponentially decreasing in time with $\Gamma$ as well.  
In other words, the effective description concerns a physical system where the number density of oscillating particles  at zero momentum monotonically decreases  with time and vanishes 
at the end of reheating.  
The relativistic matter produced by the decay process, on the other hand, can be described in terms of the perfect fluid equation of the usual form
\begin{eqnarray}\label{eqn: form3}
\dot{\rho}_r + 4H\rho_r = \Gamma\rho_{\phi} 
\end{eqnarray} 
sourced exactly by the $\Gamma\rho_{\phi}$ term.
From now on, we are reasonably assuming that the backreaction of the decay products on the system is negligible
(we aim to address modified scenarios including backreaction in a later investigation).
Therefore the physics of the reheating dynamics is described by the following well known Cauchy problem
\begin{equation}\label{eqn: reheating system_0}
\begin{cases}
\dot{\rho}_{\phi} + 3H\rho_{\phi} = - \Gamma\rho_{\phi},\quad \rho_{\phi}(t_\text{end})=\rho(\phi_\text{end})\\
\dot{\rho}_r + 4H\rho_r = \Gamma\rho_{\phi},\quad  \rho_r(t_\text{end})=0.
\end{cases}
\end{equation}
As mentioned before, $t_\text{end}$ refers to the end of inflation and coincides with the beginning of the oscillation phase, 
and the quantities $a_\text{end}$ and $H_\text{end}$ correspond, respectively, to the scale factor and the Hubble rate at the end of inflation.  
It is evident from Eq. \eqref{eqn: reheating system_0} that the initial conditions for the inflaton energy density at the beginning of the oscillation phase are given in terms of the field value $\phi_\text{end}= \phi(t_\text{end})$.  Since we are assuming that the radiation background is completely determined by the decay of the inflaton energy density for $t>t_\text{end}$, $\rho_r$ must be zero at the beginning of the inflaton decay.

\section{Solution of the Cauchy problem}

\begin{figure}[htbp]
\centering
\includegraphics[width=8.5cm, height=6cm]{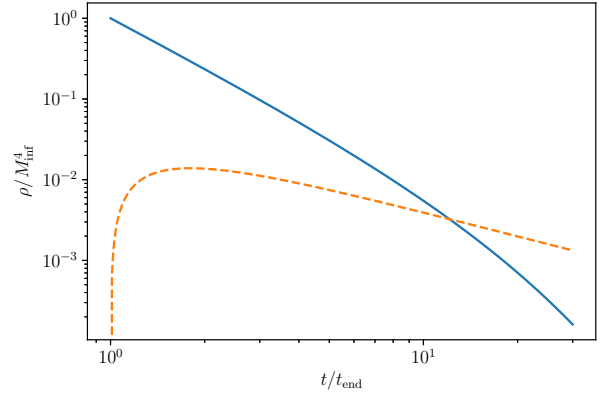}
\caption{Evolution of the inflaton energy density (solid blue line) and radiation energy density (dashed orange line) during reheating
normalized over $\rho(\phi_\text{end})\sim M^4_\text{inf}$ as function of the normalized time variable $x=t/t_\text{end}$ and with $k\sim 10^{-1}$. 
The inflaton energy density
decreases as $t^{-2}\sim a^{-3}$ up to the scale of $t=t_\text{reh}\sim \Gamma^{-1}$ where the exponential suppression starts to become dominant. 
The radiation energy density quickly increases and reaches its top value at the epoch $t_\text{max}$. 
Then it starts to decrease as $t^{-1}\sim a^{-3/2}$ and after the complete decay of the inflaton sector, the Universe become radiation dominated and radiation will follow the usual Big Bang evolution $t^{-2}\sim a^{-4}$.}
\label{fig: 1}
\end{figure}

The Cauchy problem for the evolution of the energy densities at the end of the inflationary epoch is given, as shown in the previous section, by the coupled differential equations in Eq. \eqref{eqn: reheating system_0}.
The solution of the first equation is given by
\begin{eqnarray}\label{eqn: inflaton solution}
\rho_{\phi}(t)=\rho(\phi_\text{end})\left(\frac{a_\text{end}}{a}\right)^3 e^{-\Gamma(t-t_\text{end})} \ .
\end{eqnarray}
By substituting it in the second equation and using the initial condition 
one finds \cite{27}
\begin{eqnarray}\label{eqn: radiation solution}
\frac{\rho_r(t)}{\rho(\phi_\text{end})}=\Gamma\left(\frac{a_\text{end}}{a(t)}\right)^4 \int_{t_\text{end}}^t dt'\quad \frac{a(t')}{a_\text{end}} e^{-\Gamma (t'-t_\text{end)}} .
\end{eqnarray}
Some comments are in order: if $\Gamma=0$, of course, there is not interaction with the SM particles and the energy density $\rho_{\phi}$ scales
as the one of the standard matter.  
Consequently, the inflaton field simply dilutes and the reheating phase does not occur.
The introduction of the decay width $\Gamma$, as expected, provides an interesting dynamics that depends on the ratio between the magnitude of $\Gamma$ itself and the value of the Hubble rate at the beginning of the oscillation phase 
\begin{eqnarray}
k=\frac{\Gamma}{H_\text{end}} \ ,
\end{eqnarray}
as can be argued from the exponent of $\rho_{\phi}$ in Eq.($\ref{eqn: inflaton solution}$) by roughly assuming $t_\text{end}\sim H^{-1}_\text{end}$.
If $\Gamma\sim H_\text{end}$ immediately after the end of inflation, so that $\Gamma t \gtrsim 1$, 
the energy density of the oscillations in Eq.($\ref{eqn: inflaton solution}$) is strongly suppressed and the inflaton rapidly decays into radiation.  
In this case,  reheating is an almost instantaneous process occurring at $t_\text{reh}\sim t_\text{end}$ and characterized by very efficient energy conversion and thermalization processes. Then, $\rho_r(t)$ collapses very rapidly to $\rho_r(t_\text{end}) \sim \rho(\phi_\text{end})$. Using Eq.($\ref{eqn: friedmann eq}$), one gets
\begin{eqnarray}\label{eqn: inst reh}
\rho_r(t_\text{end})\sim \rho_r(t_\text{reh})\sim \rho(\phi_\text{end})\sim 3M^2_p\Gamma^2.
\end{eqnarray}
For relativistic matter, the energy density depends on the temperature $T$ as
\begin{eqnarray}\label{eqn: static energy density}
\rho_r=\frac{\pi^2}{30}g_{E}(T)T^4 \ ,
\end{eqnarray}
where $g_E$ indicates the effective number of relativistic degrees of freedom,
\begin{equation}\label{eqn: ge}
g_{E}(T)=\sum_{b} g_{b}\left( \frac{T_b}{T}\right)^4 + \frac{7}{8}\sum_{f} g_{f}\left(\frac{T_f}{T}\right)^4, 
\end{equation}
$b$ and $f$ label bosonic and fermionic contributions, respectively.  
With $g_E\sim 100$,  Eqs.($\ref{eqn: inst reh}$) and ($\ref{eqn: static energy density}$) give rise to a reheating temperature
\begin{eqnarray}
T_\text{reh}\sim \left(\frac{90}{\pi^2 g_E}\right)^{1/4}\sqrt{\Gamma M_p}\sim 0.7 \sqrt{\Gamma M_p} \ ,
\end{eqnarray}
and subsequently, radiation will dilute as $a^{-4}$.
A more interesting situation occurs if $\Gamma\ll H_\text{end}$ (so $k\ll 1$) immediately after inflation. In this case the energy density scales like $a^{-3}$ and the energy density budget of the whole Universe is dominated by the inflaton oscillations for a long time. As a consequence, an extended reheating phase takes place  
and the cosmic scale factor, averaged over several oscillations, grows like in a matter-dominated case
\begin{eqnarray}\label{eqn: a evo}
a(t)\simeq a_\text{end}\left(\frac{t}{t_\text{end}}\right)^{2/3} \ ,
\end{eqnarray}
implying also 
\begin{eqnarray}\label{eqn: hubble rate md}
H(t)\sim \frac{2}{3t}.
\end{eqnarray}
The decay of the inflaton becomes more and more violent as the Hubble rate decreases up to the scale of $\Gamma$. Thus
\begin{eqnarray}
H(t_\text{reh})=H_\text{reh}\sim \Gamma, \quad \Gamma\sim\frac{2}{3t_\text{reh}}.
\end{eqnarray}
allows us to constrain $t_\text{reh}$, that can be naturally thought as the inflaton lifetime $\tau_{\phi}$ \cite{28}:
\begin{eqnarray}
t_\text{reh}\sim\tau_{\phi}\sim \frac{2}{3}\Gamma^{-1}.
\end{eqnarray}
Moreover, in this setting, the inflaton energy density solution in Eq.($\ref{eqn: inflaton solution}$) takes the form
\begin{eqnarray}\label{eqn: rhophix}
\rho_{\phi}(t)=\rho(\phi_\text{end})\left(\frac{t_\text{end}}{t}\right)^2 e^{-\frac{2k}{3}(t/t_\text{end}-1)}
\end{eqnarray}
while, in the limit of large inflaton lifetime/reheating duration we get
\begin{eqnarray}\label{eqn: time scales}
\Gamma t_\text{end} \sim \frac{t_\text{end}}{t_\text{reh}}\ll 1,\\
\Gamma t \sim \frac{t}{t_\text{reh}}\ll 1,\\
\Gamma(t-t_\text{end})\ll 1
\end{eqnarray}
so that $\exp(-\Gamma(t-t_\text{end}))\sim\mathcal{O}(1)$.
Then, the radiation energy density reads 
\begin{eqnarray}\label{eqn: sol1}
\frac{\rho_r(t)}{\rho(\phi_\text{end})}=\frac{2k}{3\alpha}\left(\frac{t_\text{end}}{t}\right)\left[ 1 - \left(\frac{t_\text{end}}{t}\right)^{\alpha}   \right] ,
\end{eqnarray}
where we set $\alpha=5/3$ to simplify the notation.
The evolution of the inflaton and radiation energy densities during reheating, Eq.($\ref{eqn: rhophix}$) and ($\ref{eqn: sol1}$), are plotted in Fig.($\ref{fig: 1}$) in terms of the dimensionless time variable $x=t/t_\text{end}$, with a convenient choice of parameters.

\begin{figure}[htbp]
\centering
\includegraphics[width=8.5cm, height=6cm]{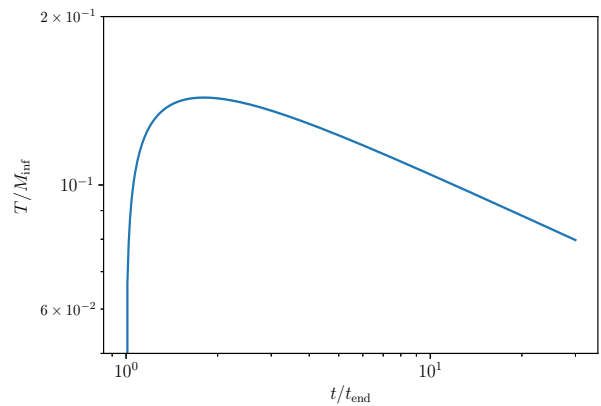}
\caption{Evolution of the temperature of the thermal bath normalized over the inflationary vacuum energy and for $k\sim 10^{-1}$.
The function follows the behavior of the radiation energy density, as one can expect.}
\label{fig: 2}
\end{figure}

The behavior of the radiation energy density function is extremely important because it allows us to estimate two crucial time scales: 
$t_\text{max}$, at which the relativistic energy density $\rho_r(t)$ reaches its maximum value, and $t_\text{reh}$, at which the reheating is completely realized or, which is the same, the inflaton has just disappeared (or is just to disappear) from the cosmological field spectrum.
By the vanishing of the derivative of $\rho_r(t)$ in Eq. ($\ref{eqn: sol1}$) one finds
\begin{eqnarray}\label{eqn: max}
\frac{t_\text{max}}{t_\text{end}}=\left(\frac{8}{3}\right)^{3/5}, \quad \mbox{ or }\quad  t_\text{max}=\left(\frac{8}{3}\right)^{3/5}t_\text{end}.
\end{eqnarray}
The height of the corresponding global maximum peak $\rho_{r}(t_\text{max})$ can be found by using Eq.($\ref{eqn: max}$) and remembering that
\begin{eqnarray}\label{eqn: simplifying rel}
\rho(\phi_\text{end})\sim 3M^2_p H^2_\text{end},\quad H_\text{end}\sim \frac{2}{3t_\text{end}}.
\end{eqnarray}
The result is \cite{29}
\begin{eqnarray}
\rho_r(t_\text{max})\sim 2\left(\frac{3}{8}\right)^{8/5} \Gamma H_\text{end} M^2_p.
\end{eqnarray}
Alternatively, $H_\text{end}$ can be connected to the inflationary scale by assuming the approximation $\rho(\phi_\text{end})\sim M^4_\text{inf}$, so that
\begin{eqnarray}
H_\text{end}\sim \frac{1}{\sqrt{3}M_p}M^2_\text{inf}
\end{eqnarray}
and
\begin{eqnarray}\label{eqn: rho max}
\rho_r(t_\text{max})\sim \frac{2}{\sqrt{3}}\left(\frac{3}{8}\right)^{8/5} M^2_\text{inf} \Gamma M_p.
\end{eqnarray}
Note, this is justified by the fact that the Hubble rate is almost constant during inflation.
The expression Eq.($\ref{eqn: rho max}$) allows us to compute the initial temperature of the produced hot matter during reheating.
In fact, using Eq.($\ref{eqn: static energy density}$) we get
\begin{eqnarray}\label{eqn: T max}
T_\text{max}&=&\left[\frac{30}{\pi^2 g_E}\frac{2}{\sqrt{3}}\left(\frac{3}{8}\right)^{8/5}\right]^{\frac{1}{4}}M^{1/2}_\text{inf}\sqrt[4]{\Gamma M_p}
\end{eqnarray}
where with $g_E \sim 100$, the prefactor reads $\sim 0.3$.
The energy density at the end of reheating $t_\text{reh}\sim \tau_{\phi}$ reads
\begin{eqnarray}\label{eqn: reheating scale}
\frac{\rho_r(t_\text{reh})}{\rho(\phi_\text{end})}=\frac{2k}{3\alpha}\left(\frac{t_\text{end}}{t_\text{reh}}\right)\left[ 1 - \left(\frac{t_\text{end}}{t_\text{reh}}\right)^{\alpha}   \right].
\end{eqnarray} 
Nevertheless, we can employ Eqs.($\ref{eqn: time scales}$) and ($\ref{eqn: simplifying rel}$) and we can use 
$t_\text{reh}\sim 2\Gamma^{-1}/3$ to find
\begin{eqnarray}\label{eqn: energy density reh}
\rho_{r}(t_\text{reh})\sim\frac{6}{5}\Gamma^2 M^2_p.
\end{eqnarray}
It is quite natural to identify the final reheating temperature of the Universe with the scale temperature 
of the relativistic plasma after thermalization. Thus, from Eqs.($\ref{eqn: static energy density}$) and ($\ref{eqn: energy density reh}$) we deduce
\begin{eqnarray}\label{eqn: T reh}
T_\text{reh}=\left(\frac{180}{5\pi^2 g_{E}}\right)^{\frac{1}{4}}\sqrt{\Gamma M_p}.
\end{eqnarray}
So, by assuming $g_E \sim 100$, the prefactor will be $\sim 0.4$. 
It is crucial to stress how the reheating  temperature depends only on the inflaton lifetime $\tau_{\phi}\sim t_\text{reh}$ and {\it not} on the scale of the inflationary vacuum energy. In other words, the larger $\Gamma$ is, the larger the final reheating temperature is and vice-versa.
Furthermore, the ratio between the maximum scale and the reheating scale can be computed and reads
\begin{eqnarray}\label{eqn:ratio1}
\frac{\rho_r(t_\text{max})}{\rho_r(t_\text{reh})}\sim p \frac{M^2_\text{inf}}{\Gamma M_p} 
\end{eqnarray}
and at the same time
\begin{eqnarray}\label{eqn:ratio2}
\frac{T_\text{max}}{T_\text{reh}}\sim p^{1/4} \frac{M^{1/2}_\text{inf}}{\sqrt[4]{\Gamma M_p}},
\end{eqnarray}
where
\begin{eqnarray}\label{eqn:constantp}
p=\frac{5}{3\sqrt{3}}\left(\frac{3}{8}\right)^{8/5}\sim 0.2, \quad p^{1/4}\sim 0.7.
\end{eqnarray}
In particular, the maximum temperature, can be rewritten in the form (see Kolb and Turner \cite{30})
\begin{eqnarray}
T_\text{max}\sim 0.4 \sqrt{ M_\text{inf} T_\text{reh}}.
\end{eqnarray}
We would like to conclude this section observing that the combination of the radiation energy density solution in Eq.($\ref{eqn: sol1}$)
and the dependence of $\rho_r$ on the temperature in Eq.($\ref{eqn: static energy density}$)
enables us to explicitly derive the temperature itself as a function of time
\begin{eqnarray}\label{eqn: temperature evolution}
\frac{T(t)}{M_\text{inf}}\sim T_0\left(\frac{t_\text{end}}{t}\right)^{1/4}\left[1-\left(\frac{t_\text{end}}{t}\right)^{\alpha}\right]^{1/4} \ .
\end{eqnarray}
The related behavior is plotted in Fig.($\ref{fig: 2}$) for a given choice of parameters, where
\begin{eqnarray}\label{eqn: T0}
T_0=\sqrt[4]{\frac{20k}{\alpha\pi^2 g_E}} \ .
\end{eqnarray}

\section{The Cauchy problem for the horizon entropy}

Another interesting aspect that deserves attention is the process of the creation and evolution of the physical entropy.
Indeed, the production of entropy accompanies the creation of relativistic matter.  
The entropy density is defined as
\begin{eqnarray}
s = \frac{2\pi^2}{45}g_{S}(T) T^3 ,
\end{eqnarray}
where $g_S$ is the number of relativistic degrees of freedom effectively contributing to it
\begin{equation}\label{eqn: gs}
g_{S}(T)=\sum_{b} g_{b}\left( \frac{T_b}{T}\right)^3 + \frac{7}{8}\sum_{f} g_{f}\left(\frac{T_f}{T}\right)^3.
\end{equation}
Using Eq.($\ref{eqn: static energy density}$), it is possible to express the entropy density also in terms of $\rho_r(t)$, getting
\begin{eqnarray}\label{eqn: entropy density radiation density}
s(t) = \frac{2\pi^2}{45}g_{S} \left(\frac{30}{\pi^2 g_E}\right)^{3/4} \rho_r^{3/4}(t) \sim 3 \rho_r^{3/4}(t).
\end{eqnarray}
Anyhow, the time dependence of the (properly normalized) entropy density will be
\begin{eqnarray}\label{eqn: energy density evolution}
\frac{s(t)}{M^3_\text{inf}}\sim s_0\left(\frac{t_\text{end}}{t}\right)^{3/4}\left[1-\left(\frac{t_\text{end}}{t}\right)^{\alpha}\right]^{3/4}, 
\end{eqnarray}
where the dimensionless coefficient $s_0$ is
\begin{eqnarray}\label{eqn: s0}
s_0=\frac{2\pi^2}{45}g_{S}\left(\frac{20k}{\alpha\pi^2 g_E}\right)^{3/4}
\end{eqnarray}

\begin{figure}[htbp]
\centering
\includegraphics[width=8.5cm, height=6cm]{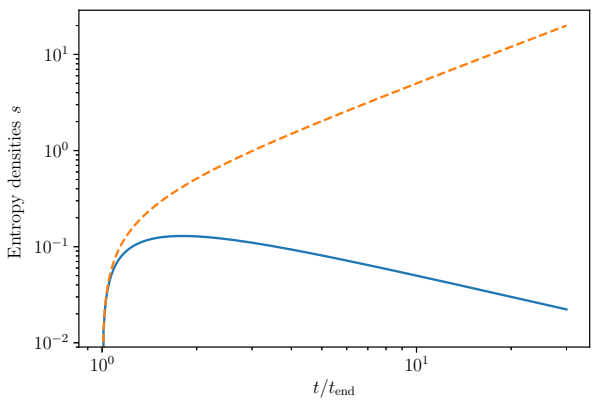}
\caption{Evolution of the normalized entropy density (solid blue curve) and of the normalized horizon entropy (dashed orange curve), for $k\sim 10^{-1}$.
The entropy density closely follows the behavior of the radiation energy density, as one can expect by definition,
while the horizon entropy grows, because the horizon patch grows as well, and takes its final value for $t\sim t_\text{reh}$.}
\label{fig: 3}
\end{figure}

However, really interesting is the evolution of the horizon entropy $S_\text{hor}$ produced during the non adiabatic reheating phase,
i.e. the entropy contained within a physical volume $\mathcal{V}_\text{hor}$ subjected to the evolution of the Universe.
Operationally, we can define the horizon entropy as
\begin{eqnarray}\label{eqn: horizon entropy}
S_\text{hor}(t)=\mathcal{V}_\text{hor}(t)s(t).
\end{eqnarray}
Although we already know the evolution of the entropy density $s(t)$,
we should suitably specify the form and the time evolution of the horizon volume.
To this end, let us suppose that the causal horizon $d_H(t_*)$, at the onset of the inflationary epoch $t_*$,
is comparable with the size of the Hubble horizon $R_H(t_*)\sim H_*^{-1}$,
so that \cite{30}
\begin{eqnarray}
d_H(t_*)\sim R_H(t_*)\sim H_*^{-1}\sim \sqrt{3}\frac{M_p}{M^2_\text{inf}}.
\end{eqnarray}
Then, the initial value of a spherical horizon volume would be
\begin{eqnarray}
\mathcal{V}_\text{hor}(t_*)\sim 4\sqrt{3}\pi \frac{M_p^3}{M^6_\text{inf}}
\end{eqnarray}
This volume violently inflates during the de Sitter-like expansion
and, it continues to grow under the milder reheating expansion.
From this point of view, the horizon volume size at some time $t>t_\text{end}$ is just
\begin{eqnarray}
\mathcal{V}_\text{hor}(t)=\left(\frac{a(t_\text{end})}{a(t_*)}\right)^3\left(\frac{a(t)}{a(t_\text{end})}\right)^3\mathcal{V}_\text{hor}(t_*)
\end{eqnarray}
that we can also write in the form
\begin{eqnarray}
\mathcal{V}_\text{hor}(t)=\left(\frac{a(t)}{a(t_\text{end})}\right)^3 \mathcal{V}_\text{hor}(t_\text{end}),
\end{eqnarray}
where the ``inflated" volume at the epoch $t_\text{end}$ comes out to be
\begin{eqnarray}
\mathcal{V}_\text{hor}(t_\text{end})=\mathcal{V}_\text{inflated}= 4\sqrt{3}\pi e^{3N} \frac{M_p^3}{M^6_\text{inf}}
\end{eqnarray}
and $N$ is the number of inflationary $e$-folds, i.e. the number of exponential expansions during inflation.
These arguments allow us to state that
\begin{eqnarray}
\frac{S_\text{hor}(t)}{\mathcal{V}_\text{inflated}}= \left(\frac{a(t)}{a(t_\text{end})}\right)^3 s(t),
\end{eqnarray}
or
\begin{eqnarray}\label{eqn: horizon entropy evolution}
\frac{S_\text{hor}(t)}{M^3_\text{inf}\mathcal{V}_\text{inflated}}=s_0 \left(\frac{t}{t_\text{end}}\right)^{5/4}\left[1-\left(\frac{t_\text{end}}{t}\right)^{\alpha}\right]^{3/4}.
\end{eqnarray}
We show the evolution of both the entropy density and the horizon entropy in Fig.($\ref{fig: 3}$).
It should be noticed that the horizon entropy grows in time because the volume of our Universe's patch grows as $a^3$.
The dominant behavior, related to the $t^{5/4}$ factor, drives the growth until the end of the reheating, 
when the entropy stabilizes at some final value, as we are going to discuss \cite{10,30}.
The described behavior of the entropy is crucial. 
Indeed, the reheating phase is the natural place where dilution of relics produced in the early universe must take place, together with baryogenesis.
In this respect, it is very interesting to get an idea about the order of magnitude of the entropy at the $t_\text{max}$ and $t_\text{reh}$ epochs.
To this end, note that Eq.($\ref{eqn: entropy density radiation density}$) provides
\begin{eqnarray}
\frac{S_\text{hor}(t)}{\mathcal{V}_\text{inflated}}\sim 3\left(\frac{a(t)}{a(t_\text{end})}\right)^3\rho_r^{3/4}(t)
\end{eqnarray}
In the limit of instantaneous reheating the evolution for times $t>t_\text{end}$ is suppressed, so it is straightforward to show that
\begin{eqnarray}
\frac{S_\text{hor}}{\mathcal{V}_\text{inflated}}\sim 7 \Gamma^{3/2}M_p^{3/2}.
\end{eqnarray}
where we used the expression of the instantaneous case for $\rho_r$.
Nevertheless,
by remembering the scale factor evolution in Eq.($\ref{eqn: a evo}$),
the relations among $t_\text{max}$ and $t_\text{end}$ in Eq.($\ref{eqn: max}$) and the amplitude of $\rho_r(t_\text{max})$ given by Eq.($\ref{eqn: rho max}$),
we get (with $g_S\sim 100$)
\begin{eqnarray}
\frac{S_\text{hor}(t_\text{max})}{\mathcal{V}_\text{inflated}}\sim 3\left(\frac{2}{\sqrt{3}}\right)^{3/4} M_\text{inf}^{3/2} \left(\Gamma M_p\right)^{3/4},
\end{eqnarray}
while the final reheating value will be
\begin{eqnarray}\label{eqn: entropy levels off}
\frac{S_\text{hor}(t_\text{reh})}{\mathcal{V}_\text{inflated}}\sim \left(\frac{6}{5}\right)^{3/4} \frac{M^4_\text{inf}}{\sqrt{\Gamma M_p}}, 
\end{eqnarray}
in such a way that 
\begin{eqnarray}\label{eqn: ratioSatone}
\frac{S_\text{hor}(t_\text{reh})}{S_\text{hor}(t_\text{max})}\sim \frac{1}{3} \left(\frac{3\sqrt{3}}{5}\right)^{3/4} \frac{M_\text{inf}^{5/2}}{\left(\Gamma M_p\right)^{5/4}}.
\end{eqnarray} 
The ratio in Eq.($\ref{eqn: ratioSatone}$) depends on the quantity $\sqrt{M_\text{inf}/\sqrt{\Gamma}}$ but at the same time is suppressed by the reduced Planck mass. 
It suggests that we cannot expect a huge entropy amplification between $t_\text{max}$ and $t_\text{reh}$.
In addition, we should note that the final horizon entropy depends on the inflationary scale because the evolution factor $\left(a(t)/a(t_\text{end})\right)^3$
does depend on it.

\section{The Cauchy problem for the generalized inflaton fluid}

In the previous sections, we have seen that in the reheating phase a simple description of the inflaton decay mechanism can be given using a perfect fluid with an Equation-of-State parameter $w$ vanishing, as is proper of the vacuum.  
However, we expect that the actual mechanism would be generically more complicated leading, for instance, to an effective description in terms of an EoS parameter that can be slightly or significantly different from the $w=0$ case.  
Scenarios of this kind are also quite naturally  emerging for instance, from String Theory. Typically, in quantum field theory (QFT), it is difficult to have an effective potential involving powers $\phi^n$ of the fields with $n > 4$. 
On the other hand, one needs to exit from a pre-existing inflationary stage characterized by $n\sim 0$
(although for a very short time). To be conservative, we may assume also $n>1$, in such a way that a reasonable range of $w$ can be 
\begin{eqnarray}
-\frac{1}{3}<w<\frac{1}{3} \ .
\end{eqnarray} 

\begin{figure}[htbp]
\centering
\includegraphics[width=8.5cm, height=6cm]{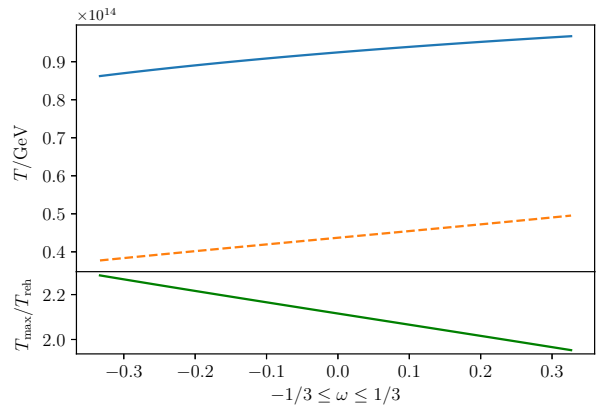}
\caption{Evolution of the maximum temperature scale (solid blue curve) and final reheating temperature (dashed orange curve) as function of the equation of state of the inflaton condensate. In the lower panel, we show the behavior of their ratio in the considered range. 
In particular, as the EoS becomes larger and larger, the ratio gets smaller and smaller.}
\label{fig: 4}
\end{figure}

Of course, the physics at very high energy scales (larger than TeV scales) is almost unknown and nothing prevents us from having scalar field configurations
that allow for larger $n$ or from additional effects mimicking analogous results for a wider range of $w$ values.
In recent years, several attempts have been done in order to analyse the effects of the EoS of the inflaton and in general of the ``reheating fluid" in specific models of inflation (see, for example, \cite{20}). 
Furthermore, the effective potential around the vacuum state could differ by
a time-independent scalar function \cite{10}.
For instance, the form of the potential could change because of peculiar  couplings of the inflaton to other bosonic and/or fermionic fields.
In this respect, the integer $\gamma$ could acquire a time dependence reflecting itself into a time dependence of the resulting EOS parameter $w=w(t)$.
In this section, however, we only consider to probe a (mean) constant value $w$ for the inflaton condensate to slightly generalize the previous, quite standard and simplified, analysis.  
A generic value of $w$ can enter the game in terms of a $\gamma=1+w$ factor within the Cauchy problem in the form
\begin{equation}\label{eqn: gen reheating system}
\begin{cases}
\dot{\rho}_{\phi} + 3H\gamma\rho_{\phi} = - \Gamma\gamma\rho_{\phi},               \quad \rho_{\phi}(t_\text{end})=\rho(\phi_\text{end})\\
\dot{\rho}_r + 4H\rho_r = \Gamma\gamma\rho_{\phi},                              \quad  \rho_r(t_\text{end})=0.
\end{cases}
\end{equation}
The source term $\gamma\Gamma\rho_{\phi}$ unavoidably implies a modified dynamics \cite{9}.
The energy density turns out to be
\begin{eqnarray}\label{eqn: gen inflaton solution}
\rho_{\phi}(t)=\rho(\phi_\text{end})\left(\frac{a_\text{end}}{a}\right)^{3\gamma}e^{-\gamma\Gamma(t - t_\text{end})}
\end{eqnarray}
while the relativistic matter density results
\begin{eqnarray}\label{eqn: gen radiation solution}
\frac{\rho_r(t)}{\rho(\phi_\text{end})}=\gamma\Gamma\left(\frac{a_\text{end}}{a(t)}\right)^4\int_{t_\text{end}}^t dt' \left(\frac{a(t')}{a_\text{end}}\right)^{4-3\gamma}e^{-\gamma\Gamma (t'-t_\text{end}}.
\end{eqnarray}
If the reheating is instantaneous, the two quantities in Eqs. \eqref{eqn: gen inflaton solution} and \eqref{eqn: gen radiation solution} obviously coincide
for $t\sim t_\text{end}$.  On the contrary, a reheating phase extended in time is governed by the evolution 
\begin{eqnarray}\label{eqn: gen scale factor}
a(t)\sim a_\text{end}\left(\frac{t}{t_\text{end}}\right)^{\frac{2}{3\gamma}}
\end{eqnarray}
with an Hubble rate 
\begin{eqnarray}\label{eqn:genHubbRa}
H(t)\sim \frac{2}{3\gamma t},
\end{eqnarray}
until the condition $H\sim \Gamma$ is satisfied. 
It happens when time is equal to
\begin{eqnarray}
t_\text{reh} \sim \frac{2}{3\gamma\Gamma} \ .
\end{eqnarray} 
Formally, the energy density of the scalar sector is given by Eq.($\ref{eqn: rhophix}$) of Sec. III.
On the other hand, in the limit of large inflaton lifetime (or long reheating phase)
\begin{eqnarray}
\gamma\Gamma t_\text{end}=\frac{t_\text{end}}{t_\text{reh}}\ll 1 \quad \gamma\Gamma t=\frac{t}{t_\text{reh}}\ll 1,
\end{eqnarray}
we get an energy density for the radiation that can be written in the same form of Eq.($\ref{eqn: sol1}$),
although now
the $\alpha$ parameter 
\begin{eqnarray}
\alpha=\frac{8-3\gamma}{3\gamma}
\end{eqnarray}
depends on $\gamma$.  
The previous equations allow us to determine the four fundamental quantities 
$\rho_r(t_\text{max})$, $T_\text{max}$, $\rho_r(t_\text{reh})$ and $T_\text{reh}$, that we are going to analyse. 
By following the same standard recipe of Sec.III to find the global peak, we vanish the time derivative of $\rho_r(t)$ that provides 
\begin{equation}\label{eqn: gen max}
\frac{t_\text{max}}{t_\text{end}}=\left(\frac{8}{3\gamma}\right)^{\frac{3\gamma}{8-3\gamma}} 
\end{equation}
telling us that the peak gets shifted at larger times as $\gamma$ or $w$ increases.
For instance
\begin{eqnarray}
\frac{t_\text{max}}{t_\text{end}}\sim
\begin{cases}
1.6,\quad w=-1/3\\
1.8,\quad w=0\\
2.0,\quad w=1/3.
\end{cases}
\end{eqnarray}
By using Eq.($\ref{eqn: gen max}$) and Eq.\eqref{eqn:genHubbRa} and with the help of
\begin{equation}\label{eqn: gen H and rho}
\rho(\phi_\text{end})\sim 3M^2_p H^2_\text{end}
\end{equation}
we find
\begin{eqnarray}
\rho_r(t_\text{max})\sim 2\left(\frac{3\gamma}{8}\right)^{\frac{8}{8-3\gamma}} \Gamma H_\text{end}M^2_p.
\end{eqnarray}
As in Sec.III, we can conveniently express it in terms of the inflationary scale
\begin{eqnarray}\label{eqn: gen rho max}
\rho_r(t_\text{max})\sim \frac{2}{\sqrt{3}}\left(\frac{3\gamma}{8}\right)^{\frac{8}{8-3\gamma}}M^2_\text{inf} \Gamma M_p, 
\end{eqnarray}
corresponding to a temperature
\begin{eqnarray}\label{eqn: gen T max}
T_\text{max}\sim \left[\frac{30}{\pi^2 g_E(T_\text{max})}\frac{2}{\sqrt{3}} \left(\frac{3\gamma}{8}\right)^{\frac{8}{8-3\gamma}} \right]^{1/4} M^{1/2}_\text{inf} \sqrt[4]{\Gamma M_p}.
\end{eqnarray}
The most important datum is, however, the reheating scale.
In the limit $t_\text{end}/t_\text{reh}\ll 1$ we find
\begin{eqnarray}\label{eqn: gen rho reh}
\rho_r(t_\text{reh})\sim \frac{6\gamma}{8-3\gamma}\Gamma^2 M^2_p \ ,
\end{eqnarray}
that generalizes the result of Eq.\eqref{eqn: energy density reh} corresponding to the case $\gamma=1$.  
The reheating temperature results in
\begin{eqnarray}\label{eqn: gen T reh}
T_\text{reh}\sim \left( \frac{180\gamma}{(8-3\gamma)\pi^2 g_E} \right)^{1/4}\sqrt{\Gamma M_p}.
\end{eqnarray}
Both the ratio of the energy densities 
\begin{eqnarray}
\frac{\rho_r(t_\text{max})}{\rho_r(t_\text{reh})}=p(\gamma)\frac{M^2_\text{inf}}{\Gamma M_p}, 
\end{eqnarray}
and the ratio of the temperature scales 
\begin{eqnarray}
\frac{T_\text{max}}{T_\text{reh}}=p^{1/4}(\gamma)\frac{M^{1/2}_\text{inf}}{\sqrt[4]{\Gamma M_p}},
\end{eqnarray}
can be given again in terms of 
\begin{eqnarray}
p(\gamma)=\frac{8-3\gamma}{3\sqrt{3}\gamma}\left(\frac{3\gamma}{8}\right)^{\frac{8}{8-3\gamma}} ,
\end{eqnarray}
that acquires a $\gamma$-dependence with respect to the one in Eq. \eqref{eqn:constantp}.
In Fig.($\ref{fig: 4}$) we show the behavior of the temperature scales as function of the EoS parameter $w$.

\begin{figure}[htbp]
\centering
\includegraphics[width=8.5cm, height=6cm]{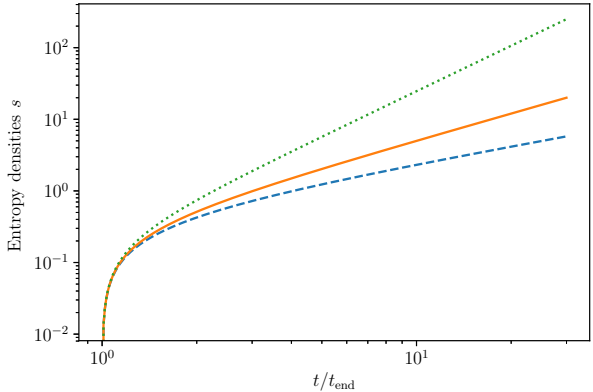}
\caption{Normalized horizon entropy for three different values of the equation of state parameter. The green dotted curve corresponds to the case $w=-1/3$,
the orange solid curve to $w=0$ while the blue dashed curve to $w=-1/3$.}
\label{fig: 5}
\end{figure}

\begin{figure}[htbp]
\centering
\includegraphics[width=8.5cm, height=6cm]{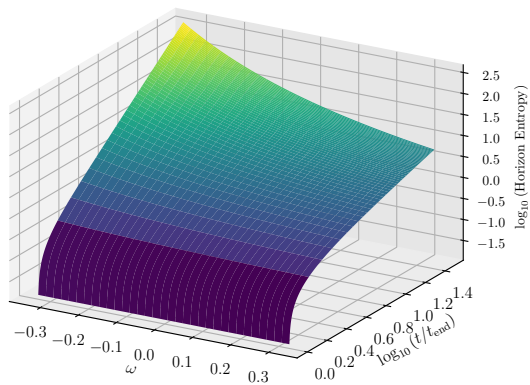}
\caption{Normalized horizon entropy as function of time and of the EoS parameter. Let us note that for $w\ge 0$ the entropy function grows almost linearly
while it grows faster and faster as the equation of state gets smaller and smaller. This justify the behavior close to $w\sim -1/3$ where the horizon
entropy grows faster than quadratically.}
\label{fig: 6}
\end{figure}

It should be noticed that the ratios can be obtained by those of Eqs. \eqref{eqn:ratio1} and \eqref{eqn:ratio2} in Sec.III by substituting the numerical factors 
with the more general prefactor $p(\gamma)$.
The behavior of the temperature and of the entropy density can be formally given by Eq.($\ref{eqn: temperature evolution}$),
($\ref{eqn: energy density evolution}$), with the proviso that, now, we have $\alpha=\alpha(\gamma)$ and therefore also 
the coefficients $T_0$ and $s_0$ depend on $\gamma$.
In particular, they turn out as increasing function of the EOS parameter $w$.
Moreover, by combining the general definition of Eq.(64) with the general cosmological scaling of Eq.(74), 
the (normalized) horizon entropy evolution can be found to be
\begin{eqnarray}
\frac{S_\text{hor}(t)}{M^3_\text{inf}\mathcal{V}_\text{inflated}}\sim s_0(\gamma) \left(\frac{t}{t_\text{end}}\right)^{\frac{2}{\gamma} - \frac{3}{4}}
\left[ 1 - \left(\frac{t_\text{end}}{t}\right)^{\alpha} \right]^{3/4}.
\end{eqnarray}
The dominant scaling behavior is given by $\sim t^{2/\gamma - 3/4}$ that means
\begin{eqnarray}
S_\text{hor}(t)\sim 
\begin{cases}
t^{9/4},\quad w=-1/3\\
t^{5/4},\quad w=0\\
t^{3/4},\quad w=1/3
\end{cases}
\end{eqnarray}
See Fig.($\ref{fig: 5}$) and Fig.($\ref{fig: 6}$) for a complete numerical result.

It is interesting to observe that the value of the horizon entropy at the epoch $t_\text{max}$
\begin{eqnarray}
\frac{S_\text{hor}(t_\text{max})}{\mathcal{V}_\text{inflated}}\sim 3 \left(\frac{2}{\sqrt{3}}\right)^{3/4} M^{3/2}_\text{inf}(\Gamma M_p)^{3/4},
\end{eqnarray}
coincides with the one of a pure matter-dominated reheating.
It means that the behavior of entropy, in an interval around $t_\text{max}$ of the decay of the inflaton coherent state, basically  does not depend upon the value of the EoS parameter $w$.
As the Universe evolution approaches the end of the reheating epoch, one has a stabilization of the entropy at a value given by
\begin{eqnarray}\label{eqn: gen entropy levels off}
\frac{S_\text{hor}(t_\text{reh})}{\mathcal{V}_\text{inflated}}\sim 3^{1-1/\gamma}\left(\frac{6\gamma}{8-3\gamma}\right)^{3/4}M_\text{inf}^{4/\gamma}(\Gamma M_p)
^{-\frac{2}{\gamma}+\frac{3}{2}}.
\end{eqnarray}
with 
\begin{eqnarray}\label{eqn:ratioShor}
\frac{S_\text{hor}(t_\text{reh})}{S_\text{hor}(t_\text{max})}=3^{-1/\gamma}\left(\frac{3\sqrt{3}\gamma}{8-3\gamma}\right)^{3/4}M^{\frac{4}{\gamma}-\frac{3}{2}}_\text{inf}
\left(\Gamma M_p\right)^{-\frac{2}{\gamma}+\frac{3}{4}} .
\end{eqnarray}
that correctly reduces to the result of Eq.\eqref{eqn: ratioSatone} as $\gamma=1$.

\section{Concluding remarks}

In this paper, we have applied a pure perturbative conversion mechanism of the energy density of the inflaton field to the energy density
of the (Standard Model) relativistic degrees of freedom in the reheating phase of the Universe, using the perfect fluid (out-of-equilibrium) formalism.  
In particular, we have extended the standard analysis of a pressure-less fluid to the case of a general EoS parameter $w\neq 0$. 
We have analysed the time evolution of the energy densities and of the physical entropy (in our inflated Universe's patch) during
the reheating stage.
The maximum energy density scale and the maximum temperature reached by the relativistic matter are those in 
Eq.($\ref{eqn: gen rho max}$) and Eq.($\ref{eqn: gen T max}$),
while the energy density and the temperature at the end of reheating appear in Eq.($\ref{eqn: gen rho reh}$) and ($\ref{eqn: gen T reh}$). 
In general, the reheating temperature depends on the source term in the equation of motion, related to the inflaton decay rate (or inflaton lifetime) and has nothing to do with the peculiar features of slow roll dynamics, for instance with the vacuum energy that drives inflation, $M_\text{inf}$, or with the scalar field excursion. 
On the contrary, the detailed particle production by inflaton decay depends upon the inflaton potential.
The reheating temperature, Eq.($\ref{eqn: gen T reh}$), 
determines the energy scale at which the thermalization process, due to continuous scatterings within the plasma of the relativistic matter, provides the conditions for the grateful exit, namely for the beginning of the standard Friedmann radiation dominated era. 
However, such energy scale may be sensitive to many perturbative quantum field theory effects.
For example, the reheating temperature would be very sensitive to the full spectrum of inflaton decay products \cite{31},
or could be sensitive to the plasma masses acquired by the inflaton decay products \cite{32}. 
Furthermore,
it has been shown that thermalization can also end much later than the completion of the inflaton decay and the beginning of the radiation dominance. 
In this case, the ``effective" reheating scale $T_\text{reh}^\text{eff}$, could fall well within the radiation epoch and, consequently, could be
much lower than the standard prediction in Eq.($\ref{eqn: gen T reh}$), with $T_\text{reh}^\text{eff}\ll T_\text{reh}$\cite{33}.
More recently, it has been discussed that the whole process could involve more cosmological sectors that can influence the final reheating scale \cite {34}.
Of course, the perturbative approach to reheating is partially incomplete and 
inherently characterized by some limitations \cite{18}. 
To describe the details of the particle production and more realistic results, one is indeed forced to resort to non-perturbative 
effects like, for instance, when very large couplings \cite{11,12,13,14,15,16,17,18} are present.

\begin{acknowledgments}
This work was supported in part by the Ministero
dell' Istruzione, dell' Universita' e della Ricerca (MIUR) - Progetti di Rilevante Interesse Nazionale (PRIN) 
contract 2015MP2CX4002, ``Non perturbative Aspects of Gauge Theories and Strings".
A.D.M would like to thank J.Martin and Y.Shtanov for useful private communications.
P.C would like to thank the company ``L'isola che non c'\`e S.r.l" for the support.
\end{acknowledgments}


\nocite{*}

\bibliography{apssamp}

\begin{thebibliography}{10}
\bibitem{27} Here we use the $'$ to distinguish the integration variable from the upper limit of integration.
\end{thebibliography}

\end{document}